\documentclass[12pt]{article}
\usepackage{graphicx}

\def\be{\begin{equation}}
\def\ee{\end{equation}}
\def\beq{\begin{eqnarray}}
\def\eeq{\end{eqnarray}}

\begin{document}
\title{\bf 
Asymptotic Behavior of the Emptiness Formation Probability
in the Critical Phase of   XXZ Spin Chain
}

\author{
        V. E.  Korepin \\
{\it C.N.~Yang Institute for Theoretical Physics}\\
{\it State University of New York at Stony Brook}\\
{\it Stony Brook, NY 11794--3840, USA}\\
\\ S. Lukyanov \\
{\it Department of Physics and Astronomy, Rutgers University}\\
{\it Piscataway, NJ 08855-0849, USA}\\
\\     Y. Nishiyama \\
{\it Department of Physics, Faculty of Science, Okayama University}\\
{\it Okayama 700-8530, Japan}\\
\\     M. Shiroishi \\ 
{\it Institute for Solid State Physics, University of Tokyo}\\
{\it Kashiwanoha 5-1-5, Kashiwa, Chiba, 277-8571, Japan}
}

\maketitle\thispagestyle{empty}
\abstract{
We study the Emptiness Formation Probability (EFP) for 
the spin 1/2  $XXZ$  spin chain. EFP $P(n)$ detects a formation 
of ferromagnetic string of the length $n$ in the ground state.
It is expected that EFP decays in a Gaussian way for large strings 
$P(n)\sim n^{-\gamma} C^{-n^2}$. Here, we propose the explicit 
expressions for the rate of Gaussian decay  ${C}$ as well as for 
the exponent $\gamma$. In order to confirm the validity of our 
formulas, we employed an {\it ab initio} simulation technique of 
the density-matrix renormalization group to simulate $XXZ$ spin 
chain of sufficient length. Furthermore, we performed Monte-Carlo 
integration of the Jimbo-Miwa multiple integral for ${P(n)}$.
Those numerical results for ${P(n)}$ support our formulas 
fairly definitely. 
}

\newpage
\vspace*{1cm}
\noindent
%\section{}
%\setcounter{equation}{0}
%\renewcommand{\theequation}{1.\arabic{equation}}
We consider interacting spin 1/2  on the one-dimensional 
infinite lattice.
The model is described by the 
Hamiltonian
\begin{equation}
{\cal H}=\sum_{j=-\infty}^{\infty} 
\left\{ S^x_j S^x_{j+1} + S^y_j S^y_{j+1} 
+ \Delta \left(S^z_j S^z_{j+1} -\frac{1}{4} \right) \right\}, \label{XXZ}
\end{equation}
where  $S= \sigma /2$ and  $\sigma$ are Pauli matrices.
It  exhibits diverse physics with varying the anisotropy parameter $\Delta$.
For $\Delta \le -1$ the ground state  has a long-range ferromagnetic 
 order, 
and a finite excitation gap  opens in the low-lying spectrum. 
For $\Delta > 1$ the ground 
state  develops  antiferromagnetic order, the gap also exists.
For moderate anisotropy $-1 < \Delta \le 1$, the
antiferromagnetic order
is dissolved by quantum  fluctuations.
The  gap closes,  and
the long-range asymptotic of
correlation function  
$\langle S_i S_j \rangle$
decays as a power   law   at
zero temperature.

In this article  we study  the  Emptiness Formation Probability (EFP),
\begin{equation}
P(n) = \Big\langle \prod_{j=1}^n \left(S_j^z + \frac{1}{2} \right)
\Big\rangle, 
\label{EFP}
\end{equation}
in the domain of critical phase.
The importance of EFP was emphasized in the book \cite{BIK93}. 
It reflects the nature of the ground-state
and it is a good indicator of the phase-separation.

In spite of its seemingly complicated definition,  EFP has emerged
quite naturally in the course of studies of the quantum integrability 
of the $XXZ$ spin chain.
The first  progress in calculating of EFP was achieved
in Ref.~\cite{Takahashi77},  where $P(3)$ was obtained  for
the isotropic antiferromagnetic chain $(\Delta=1)$.
Subsequently, Jimbo and Miwa derived multiple integral representation 
for all correlation functions by means of vertex operators approach
\cite{Jimbo95,Jimbo96}. 
Their integral  formula for EFP is:
\begin{eqnarray}
P(n) &=& \left(- \nu \right)^{-\frac{n(n+1)}{2}} \int_{-\infty}^{\infty} 
\frac{{\rm d} \lambda_1}{2 \pi} \cdots \int_{-\infty}^{\infty} 
\frac{{\rm d} \lambda_n}{2 \pi}  \prod_{a>b} \frac{\sinh (\lambda_a - \lambda_b)}
{\sinh \left( \left(\lambda_a-\lambda_b - {\rm i} \pi \right) \nu \right)}  \nonumber \\
& & \hspace{1cm} \times \prod_{k=1}^{n} \frac{\sinh^{n-k} \left( \left( \lambda_k + {\rm i} \pi/2 
\right) \nu \right) \sinh^{k-1} \left( \left(\lambda_k - {\rm i} \pi/2 \right) 
\nu \right)}{\cosh^n (\lambda_k)}, 
\label{Jimbo-Miwa}
\end{eqnarray}
where the  parameter ${\nu}$ is related to the anisotropy $\Delta$  as
\begin{equation}
\nu = \frac{1}{\pi} \cos^{-1}(\Delta). \label{nu}
\end{equation}
%Later Miwa and Jimbo also noticed that  correlation
% functions 
%in the gap-less regime  satisfy quantum Knizhnik-Zamolodchikov equation.
%This provide the best way 
%of evaluation of correlation functions, see \cite{Boos02b}.
The integral representation 
was reproduced   in the framework
of  algebraic Bethe ansatz in \cite{Kitanine00}. 
A good deal of new developments have been reported for EFP. 
For example, in the case of  ${\Delta=1/2}$ the simple 
formula for $P(n)$ has been conjectured in \cite{Razumov00} and  
then proved in \cite{Kitanine02}. For ${\Delta=0}$ the asymptotic 
form of $P(n)$ as $n\gg 1$ was found  in \cite{Shiroishi01}.
In either  case the large-$n$  asymptotic behavior
is given by
$ P(n) \sim n^{-\gamma}\, C^{- n^2}
$
with 
\begin{eqnarray}
C &=& \sqrt{2}, \ \ \ \ \  \gamma = \frac{1}{4}, 
\ \ \ \ \ (\nu = 1/2),  \nonumber \\ 
C &=&  \frac{8}{3\sqrt{3}}, \ \ \ \gamma = \frac{5}{36}, \ \ \ \ (\nu=1/3). 
\label{exact}
\end{eqnarray}
Moreover, ${P(n)}$ was calculated recently for $\Delta=1$ for strings of the length $n\le6$,
see papers
 \cite{Boos01,Boos02,Boos02b}. 
Their result suggests that ${P(n)}$ at $\Delta=1$ decays 
in a Gaussian way as well.
Meanwhile, the asymptotic behavior of EFP for the whole critical regime 
was analyzed in   field-theoretical 
framework, by Abanov and Korepin \cite{Abanov02}. The Gaussian decay naturally appears in this approach. 

Based on those developments, it would appear 
reasonable\ that   the following  asymptotic
form of EFP  
\begin{equation}
P(n) \simeq A\  n^{-\gamma}\,  C^{- n^2}, \label{Asym} 
\end{equation} 
holds for all over 
the critical regime ${(-1 < \Delta \le 1)}$. 
         We propose the explicit expressions for the rate of
 Gaussian decay  ${C}$ and the power-law exponent 
$\gamma$:
\begin{eqnarray}
C &=& \frac{\Gamma^2(1/4)}{\pi \sqrt{2 \pi}} \exp \left\{ - \int_0^{\infty} 
\frac{{\rm d} t}{t} 
\frac{\sinh^2(t \nu) {\rm e}^{-t}}{\cosh(2 t \nu) \sinh(t)} \right\}, 
\label{C} \\
\gamma &=& \frac{1}{12} + \frac{\nu^2}{3(1-\nu) }.  \label{gamma}
\end{eqnarray}
One can confirm that above  formulas reproduce the
exact results for $\Delta =0$ and $\Delta =1/2$\ (\ref{exact}). 

In order to confirm the validity of the formulas for general anisotropies 
${0 \le \nu  < 1 }$, we have performed extensive numerical calculations 
of two kinds:
One is the first-principle simulation method, namely,
the density-matrix renormalization group (DMRG) 
\cite{White92,White93,Peschel99},
and the other is the Monte-Carlo numerical integration of
the multiple integration formula (\ref{Jimbo-Miwa}) of Jimbo and Miwa.
As noted afterwards, those methods are compensative, and we were
able to
perform reliable simulations for over all the critical regime.
As a result, we could confirm fairly definitely that
the above general formulas are indeed correct.

Let us turn to addressing the numerical-simulation results.
In Tables 1-4, we have listed the DMRG results of EFP for $\nu =0.2,...,0.8$.
(The methodological details will be explained afterwards.)

In addition, in each table, we have presented the logarithm of the ratio 
of two adjacent EFPs, which should behave in the form 
\begin{equation}
\ln \left(\frac{P(n)}{P(n+1)}\right) \simeq \gamma \ln (1+\frac{1}{n})
  + (2n+1) \ln C, \hspace{1.5cm} \label{Asym2}
\end{equation}
according to the long-distance asymptotic formula (\ref{Asym}).
Note that, after taking the ratio, we are able to kill the contribution
of the constant factor $A$ of (\ref{Asym}), for which, at present, 
we have no analytical prediction. 
(We, however, could estimate ${A}$ numerically from our DMRG data 
as is shown in Table 5.) 
%We have numerical values for$A$ \footnote{For XXX antiferromagnet 
%$A=0.841$.}.  
In this way, the resultant processed data can be directly comparable with 
the analytical conjecture.

We see that our processed DMRG data $\ln(P(n)/P(n+1))$ are extremely
close to the analytical prediction (\ref{Asym2}). Actually we find 
that they coincide up to two or three digits 
for general values of $n$ and $\nu$.
Therefore we conclude that
our formulas for the asymptotic form are valid indeed
for all over the regime ${0 \le \nu <1}$. 
Especially see the Table 6, where 
we compare the asymptotic formula with the known exact values for ${\Delta=1}$ 
\cite{Boos01,Boos02,Boos02b}.  Note that, in this case, we have 
${C = \Gamma^2(1/4)/(\pi \sqrt{2 \pi}) = 1.66925..., \ \ \gamma = 1/12}$,
and ${A = 0.841.}$

We, however, remark that rather large discrepancies are seen for such cases 
either very large ${n}$ for small ${\nu}$ or
small $n$ for large $\nu$. The former deviations are merely due to the 
numerical round off errors. Since in computers, real numbers are stored 
in 8-byte space, and the precision is of the order of $10^{-15}$ at best.
Hence it is in principle difficult to calculate the correlations less
than the magnitude $< 10^{-12}$ reliably. The latter deviations are 
not so surprising, because our general formulas should be justified for 
long distances of EFP. This short-range deviation will be further exploited 
in the succeeding Monte-Carlo numerical integration analyses.

\begin{table}[htpb]
\begin{center}
\caption{DMRG data for ${P(n)}$ : ${\nu=0.2}$}
\label{table1}
\begin{tabular}{@{\hspace{\tabcolsep}\extracolsep{\fill}}cccc} \hline
$n$ & ${P(n)}$  & ${\ln(P(n)/P(n+1))}$ & Asymptotics (\ref{Asym2}) \\ \hline
1  &  ${0.5}$ \ &   1.505 \ 
   &   1.512                   \\ 
2  &  ${1.111 \times 10^{-1}}$ \ &   2.447  \ 
   &   2.447                          \\ 
3  &  ${9.614 \times 10^{-3}}$ \ &   3.397 \ 
   &   3.397                           \\ 
4  &  ${3.218 \times 10^{-4}}$ \ &   4.353 \ 
   &   4.353                            \\ 
5  &  ${4.140 \times 10^{-6}}$ \ &   5.312 \ 
   &    5.312                            \\   
6  &  ${2.041 \times 10^{-8}}$ \ &   6.276  \
   &   6.271                           \\  
7  &  ${3.837 \times 10^{-11}}$ \ &   7.187 \
   &   7.232                          \\  
8  &  ${2.901 \times 10^{-14}}$ \ &   4.683 \
   &   8.192                           \\  
9  &  ${2.682 \times 10^{-16}}$ \ &   3.923 \
   &     9.155                            \\ \hline
\end{tabular}
\end{center}
\end{table}

\begin{table}[htpb]
\begin{center}
\caption{DMRG data for ${P(n)}$ : ${\nu=0.4}$}
%\label{table1}
\begin{tabular}{@{\hspace{\tabcolsep}\extracolsep{\fill}}cccc} \hline
$n$ & ${P(n)}$  & ${\ln(P(n)/P(n+1))}$ & Asymptotics (\ref{Asym2}) \\ \hline
1  &  ${0.5}$ \ &    1.319  \ 
   &    1.320                   \\ 
2  &  ${1.337 \times 10^{-1}}$ \ &  2.070   \ 
   &    2.071                          \\ 
3  &  ${1.687 \times 10^{-2}}$ \ &  2.851 \ 
   &   2.851                           \\ 
4  &  ${9.752 \times 10^{-4}}$ \ &  3.640 \      
   &  3.640                         \\ 
5  &  ${2.561 \times 10^{-5}}$ \ &   4.434 \           
   &  4.433                             \\   
6  &  ${3.039 \times 10^{-7}}$ \ &   5.231  \          
   &   5.230                          \\  
7  &  ${1.626 \times 10^{-9}}$ \ &   6.026 \           
   &   6.025                           \\  
8  &  ${3.924 \times 10^{-12}}$ \ &   5.862 \  
   &   6.823                            \\  
9  &  ${1.117 \times 10^{-14}}$ \ &   2.824 \   
   &   7.621                            \\ \hline
\end{tabular}
\end{center}
\end{table}

\begin{table}[htpb]
\begin{center}
\caption{DMRG data for ${P(n)}$ : ${\nu=0.6}$}
% \label{table1}
\begin{tabular}{@{\hspace{\tabcolsep}\extracolsep{\fill}}cccc} \hline
$n$ & ${P(n)}$  & ${\ln(P(n)/P(n+1))}$ & Asymptotics (\ref{Asym2}) \\ \hline
1  &  ${0.5}$ \ &   1.106 \ 
   &   1.125                   \\ 
2  &  ${1.655 \times 10^{-1}}$ \ &   1.594  \ 
   &   1.588                          \\ 
3  &  ${3.360 \times 10^{-2}}$ \ &   2.121 \ 
   &   2.115                           \\ 
4  &  ${4.028 \times 10^{-3}}$ \ &   2.667 \ 
   &   2.663                            \\ 
5  &  ${2.798 \times 10^{-4}}$ \ &   3.223 \ 
   &   3.221                            \\   
6  &  ${1.115 \times 10^{-5}}$ \ &   3.784  \
   &   3.783                           \\  
7  &  ${2.534 \times 10^{-7}}$ \ &   4.348 \
   &   4.348                          \\  
8  &  ${3.278 \times 10^{-9}}$ \ &   4.909 \
   &   4.915                           \\  
9  &  ${2.420 \times 10^{-11}}$ \ &  5.231 \
   &   5.483                            \\ \hline
\end{tabular}
\end{center}
\end{table}

\begin{table}[htpb]
\begin{center}
\caption{DMRG data for ${P(n)}$ : ${\nu=0.8}$}
% \label{table1}
\begin{tabular}{@{\hspace{\tabcolsep}\extracolsep{\fill}}cccc} \hline
$n$ & ${P(n)}$  & ${\ln(P(n)/P(n+1))}$ & Asymptotics (\ref{Asym2}) \\ \hline
1  &  ${0.5}$ \ &   0.896 \ 
   &   1.250                   \\ 
2  &  ${2.041 \times 10^{-1}}$ \ &   1.120  \ 
   &   1.221                          \\ 
3  &  ${6.663 \times 10^{-2}}$ \ &   1.361 \ 
   &   1.387                           \\ 
4  &  ${1.709 \times 10^{-2}}$ \ &   1.615 \ 
   &   1.615                            \\ 
5  &  ${3.400 \times 10^{-3}}$ \ &   1.879 \ 
   &   1.870                            \\   
6  &  ${5.196 \times 10^{-4}}$ \ &   2.150  \
   &   2.139                           \\  
7  &  ${6.050 \times 10^{-5}}$ \ &  2.430 \
   &   2.417                          \\  
8  &  ${5.336 \times 10^{-6}}$ \ &   2.708 \
   &   2.701                           \\  
9  &  ${3.549 \times 10^{-7}}$ \ &   2.996 \
   &    2.988                           \\ \hline
\end{tabular}
\end{center}
\end{table}

\begin{table}[htpb]
\begin{center}
\caption{Numerical estimation of the factor ${A}$ }
% \label{table1}
\begin{tabular}{@{\hspace{\tabcolsep}\extracolsep{\fill}}crcll} \hline
${\nu}$ & ${\Delta}$  & ${C}$ & ${\gamma}$ & ${A}$   \\ \hline
0.0 & 1.0       \ & 1.66925    & \ 0.0833   & 0.841   \\ 
0.2 & 0.8090    \ & 1.61803    & \ 0.1      & 0.816   \\ 
0.4 & 0.3090    \ & 1.49207    & \ 0.1722   & 0.747   \\ 
0.6 &  -0.3090  \ &  1.33168   & \ 0.3833   & 0.68    \\ 
0.8 &  -0.8090  \ &  1.16287   & \   1.15   & 0.9      \\ \hline
\end{tabular}
\end{center}
\end{table}

\begin{table}[htpb]
\begin{center}
\caption{Comparison with exact values for ${P(n)}$ at ${\Delta=1}$}
% \label{table1}
\begin{tabular}{@{\hspace{\tabcolsep}\extracolsep{\fill}}cccc} \hline
$n$ & ${P(n)}$  & ${\ln(P(n)/P(n+1))}$ & Asymptotics (\ref{Asym2}) \\ \hline
1  &  ${0.5}$ \ &   1.58685 \ 
   &   1.59489                   \\ 
2  &  ${1.02284 \times 10^{-1}}$ \ &   2.59643  \ 
   &   2.59566                          \\ 
3  &  ${7.62415  \times 10^{-3}}$ \ &  3.60989 \ 
   &   3.61060                           \\ 
4  &  ${2.06270 \times 10^{-4}}$ \ &   4.63019 \ 
   &   4.62998                            \\ 
5  &  ${2.01172 \times 10^{-6}}$ \ &   5.65115 \ 
   &   5.65133                            \\   
6  &  ${7.06812 \times 10^{-9}}$ \ &     \
   &                              \\ \hline
\end{tabular}
\end{center}
\end{table}

So far, we had analyzed processed EFP data.
In Fig. 1, we present the raw (unprocessed) EFP data obtained with 
the DMRG method.
The dotted lines are the analytical conjecture with the factors given 
in Table 5.
We see that our first-principle data are well-fitted by
our general formulas (\ref{C}) and (\ref{gamma}) for various values of $\nu$.
However, as noted above, for some cases, there appear deviations.
\begin{figure}[htbp]
\begin{center} 
\includegraphics{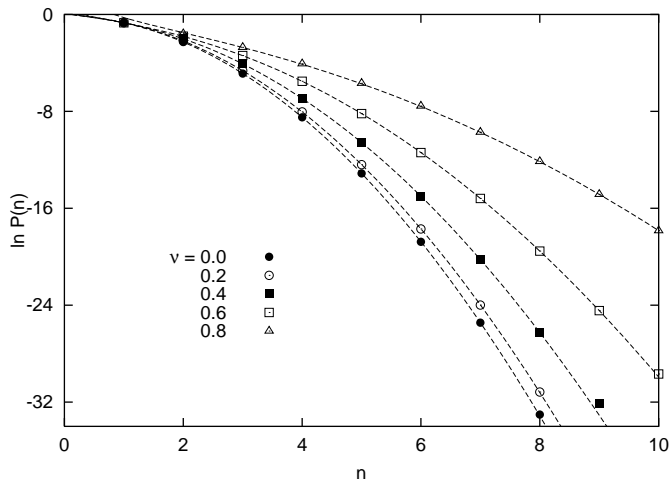}
\end{center}
\caption{EFP is plotted for various valued of distances $n$ 
and anisotropies $\nu$. 
We have employed the DMRG method.
The dotted lines are our conjecture
(\ref{Asym}) with the formulas (\ref{C}) and (\ref{gamma}).} 
\label{DMRG}
\end{figure}

Let us turn to addressing details of the DMRG method
\cite{White92,White93,Peschel99}.
DMRG is a sort of computer-aided real-space decimation,
where the number of states for a renormalized block 
is retained within a bound tractable in computers. 
In this way, through the successive applications of DMRG, 
we are able to access very large system sizes.
Hence, this method is suitable for surveying the long-distance
asymptotic behavior of EFP.
As a matter of fact, 
efficiency of the method was demonstrated in the preceding
works
\cite{Shiroishi01,Boos02}:
This success was rather unexpected, because
in general, DMRG is not very efficient for such systems 
exhibiting criticality.
In particular, long-distance asymptotic behavior of two-point
correlation function is deteriorated rather severely.
Those pathologies would be due to the truncation of bases through
numerical renormalizations.
However, as far as EFP is concerned, DMRG is proven to be free from such 
difficulties.
Below, we will overview technical points of our numerical simulation.
Our algorithm is standard \cite{White92,White93}, 
and we refer readers to a recent proceeding
\cite{Peschel99}
for full account of methodological details.
We have employed the so-called infinite algorithm, which is adequate
to investigate bulk properties at the ground state.
(For the purpose of studying finite-size scaling behavior, 
the finite algorithm would be more suitable.)
We have remained, at most, $m=300$ bases for a renormalized block.
The density matrix eigenvalue $\{w_\alpha\}$ of 
remaining bases indicates the statistical weight of each remained state.
We found $w_\alpha > 5 \cdot 10^{-12}$:
That is, we have remained almost all relevant states
with appreciable statistical weight $w_\alpha > 5 \cdot 10^{-12}$,
which may indicate error of the present simulation.
We have repeated 300 renormalizations, and hence, total system size
extends to $L=600$.
The DMRG data alternates in turn through renormalizations:
Note that the number of spins consisting a renormalized block increases
by one after another through renormalizations.
The problem is that the Hilbert-space structures are incompatible
with respect to those cases whether the block contains even or odd spins.
(Haldane system ($S=1$) is not affected by this difficulty.)
Therefore, we have taken arithmetic mean over those two cases.

The DMRG method is not efficient in the close vicinity of the 
ferromagnetic isotropic point $\nu=1$.
Because in the vicinity of that point, the spin-wave velocity tends to vanish, 
and there appear numerous nearly-degenerated
low-lying levels.
Those nearly-degenerated levels are very hard to resolve in the
process of numerical diagonalization.
The diagonalization is a significant part in the DMRG procedure,
and hence, DMRG becomes hardly applicable.
In order to compensate this drawback, we have employed another
numerical method, that is, the Monte-Carlo integration.
We will explain it in the following.

\begin{figure}[htbp]
\begin{center} 
\includegraphics{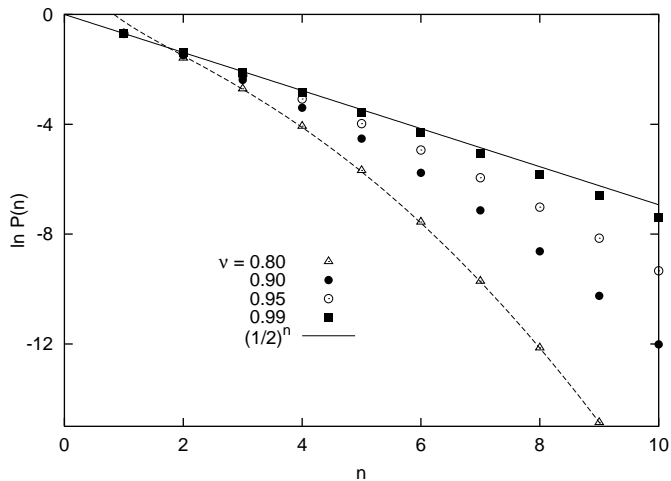}
\end{center}
\caption{EFP is plotted for various valued of distances $n$ 
and anisotropies $\nu$. 
We have employed the Monte-Carlo integration method
for Eq. (\ref{Jimbo-Miwa}).
The dotted line is Eq. (\ref{Asym}) for ${\nu=0.8}$.
} 
\label{EFP_fig} 
\end{figure}

In Fig. 2, we have presented numerical data with Monte-Carlo integration
for the multiple integral formula (\ref{Jimbo-Miwa}) of Jimbo and Miwa.
The technical details will be explained afterwards.
 From the plot, we see that the decay of EFP is gradually 
modified as for $\nu \to 1$, and surprisingly enough, 
the decay approaches the simple exponential formula of $(1/2)^n$
eventually.
This feature is precisely in accord with the aforementioned finding that 
DMRG data deviate from the Gaussian formula as for $\nu \to 1$.
Therefore, we see that at least for short-distance asymptotic form
of EFP is governed by the simple exponential decay $0.5^n$
as $\nu \to 1$.
As a matter of fact, right at $\nu=1$, one may easily verify
the pure exponential decay
$P(n)$ from the Jimbo-Miwa formula.

The crossover from Gaussian to exponential decay suggests that
the quantum fluctuations are suppressed as $\nu \to 1$.
As as matter of fact, field theoretical consideration \cite{Abanov02}
reveled that EFP measures the probability of formations of
ferromagnetic islands with size $n$ surrounded
by antiferromagnetic background from the viewpoint of
Euclidean space-time.
Therefore, our result suggests that this Euclidean space-time
picture is deteriorated for $\nu \to 1$, because
the spin-wave velocity tends to vanish, and thus, the size of the islands
grows along the imaginary-time direction abruptly.

In the following, we explain technical details of the Monte-Carlo 
integration.
We have used the subroutine package described in the textbook \cite{NRF}.
As a random number generator, we have employed ``Mersenne Twistor", 
MT19937 in  Ref.~\cite{Matsumoto98}. 
For each plot, we have performed eight-billion Monte-Carlo steps.
This main Monte-Carlo procedure is preceded by preliminary five million
Monte-Carlo steps which are aimed to improve the efficiency
of Monte-Carlo sampling
by surveying the integrand in the multi-dimensional
space.
Irrespective of $n$ and $\nu$, we found that the statistical errors
are of the order of $10^{-8}$.

To summarize, we have advocated compact explicit formulas
(\ref{C}) and (\ref{gamma}) for the long-distance asymptotic form behavior 
of EFP.
The formulas reproduce the presently-available exact
 results obtained at special solvable points
of $\Delta=0$ and $1/2$, and therefore, we expect that our formulas are
valid for over all the critical regime.
We have performed extensive simulations of DMRG and Monte-Carlo
integration.
As a result, we found that our general formulas are
indeed correct over the critical regime; see Tables and Fig. 1.
In other words, EFP decays obeying the Gaussian formula (\ref{Asym})
in the domain of criticality,
and the decay rate becomes slower in the ferromagnetic side.
This tendency is understood by a physical picture that EFP detects
a formation of condensed particles (ferromagnetic string).
Apparently, for ferromagnetic side, the condensation becomes promoted,
and EFP gets enhanced, albeit the asymptotic form is still maintained to be
Gaussian.
However, as for the extreme limit to the ferromagnetic isotropic point, 
the Gaussian decay is smeared out by an exponential form
at least for short distances.
This fact reflects the suppression of quantum fluctuations in the ferromagnetic
side.
In other words, this crossover may be regarded as the precursor of the onset
of true (long-range order) phase-separation for $\Delta<-1$, 
where EFP should not decay, but may saturate to a certain finite resident value.
Finally we like to remark that the rate of Gaussian decay ${C}$, Eq.(\ref{C}), can be 
evaluated analytically for some other rational values of ${\nu}$. For example, 
we have found 
\begin{eqnarray}
C|_{\nu=1/4} &=& 2^{\frac{1}{4}} {\rm e}^{\frac{\rm G}{\pi}}, \ \ \ \ \ \ 
({\rm G : Catalan \ number}),  \nonumber \\
C|_{\nu=1/6} &=& 2 \sqrt{\frac{2}{3}}, \ \ \ \ 
C|_{\nu=2/3} = \frac{16(\sqrt{2}-1)}{3 \sqrt{3}}, 
\end{eqnarray}
by use of the integral representations for the logarithm of (multiple) Gamma functions. 
These facts may indicate a possibility to prove our analytic formulas (\ref{Asym}) 
rigorously for these values of ${\nu}$ just in the similar way as ${\nu=1/2}$ and ${1/3}$. 

{\bf Acknowledgment}
We thank A. Abanov, H.Boos, F.Essler, F.Smirnov, M.Takahashi and A. Zamolodchikov  for
useful discussions. V. Korepin  was supported by by  NSF Grant PHY-9988566, 
S. Lukyanov was supported in part  by DOE Grant DE-FG02-96ER40959,
Y. Nishiyama was supported by Grant-in-Aid for Young Scientists No. 13740240,
M. Shiroishi was supported by Grant-in-Aid for Young Scientists No. 14740228.
%V. Korepin is grateful to the organizers of the Conference on Random Matrix
% Theory and Combinatorics
%[Courant Institute of Mathematical Sciences, New York University, June 2-6, 2002] for the opportunity to
% present the results.

\end{document}